# Quasi-Degenerate Neutrino Masses in Terms of Mass-Squared Differences

E. M. Lipmanov

40 Wallingford Road #272, Brighton MA 02135

Abstract

The absolute neutrino masses are obtained in terms of the atmospheric and solar mass-squared differences within the framework of low energy phenomenology by suggestion of a quantitative analogy between the hierarchies of the CL and neutrino mass ratios. It points to a Q-D three neutrino mass pattern with the neutrino mass scale $m_\nu \cong \Delta m^2_{atm}/\sqrt{(2\sqrt{2}\Delta m^2_{sol})}$ likely located in the range 0.1 – 0.3 eV, and the best-fit value $m_\nu \cong$ 0.18 - 0.20 eV. Restrictions on the neutrino mass scale from the WMAP data are considered. Possible indirect evidence in favor of Q-D neutrino masses is noted.

## 1. Introduction

The known sharp contrast between the neutrinos and CL is the very large difference of their mass scales. The CL masses $m_e$, $m_\mu$ and $m_\tau$ are well known [1]. Two large mass ratios and a large hierarchy of these mass ratios characterize the mass pattern of the CL:

$$m_\mu/m_e \gg 1, \quad m_\tau/m_\mu \gg 1, \quad (m_\tau/m_\mu)^2 \cong (m_\mu/m_e)\sqrt{2}. \qquad (1)$$

The discovery of the finite neutrino masses in the neutrino oscillation experiments [2,3,4] does raise the question: what is the neutrino mass pattern and what relation is there between the two mass patterns if any? This problem is widely discussed [5] in the contexts of different basic extensions of the SM with higher mass and energy scales. There is no definite answer to this question since the exact absolute values of the neutrino masses are unknown as yet, while the neutrino oscillation data give only neutrino mass-squared differences.

In this note, an attempt is made to answer the above question in the framework of low energy phenomenology guided by the neutrino oscillation data against the background of a virtual broken lepton mass eigenstate symmetry (flavor problem). In spite of the disparity of the mass scales, an analogy between the neutrino and CL mass ratio hierarchies[1] is suggested and described by an extension of the condition (1), taking into consideration the factual violation of the lepton mass state symmetry. This analogy relates the three absolute neutrino masses to the two oscillation mass-squared differences.

---

[1] For short, the term "mass hierarchy" will be used in place of a more accurate phrase "nonlinear hierarchy of the dimensionless deviations from mass-degeneracy".

## 2. Two opposite lepton mass ratio patterns

By definition, the sequence of the lepton masses (CL or neutrinos) let be
$$m_1 < m_2 < m_3. \tag{2}$$
In view of two basic experimental facts — the CL mass ratio hierarchy in (1) and the hierarchy of the atmospheric and solar neutrino mass-squared differences — we suggest an approximate unifying nonlinear relation of the lepton mass ratios $x_2 \equiv m_3/m_2$ and $x_1 \equiv m_2/m_1$ at a common low scale,
$$(x_2 - 1)^2 \cong (x_1 - 1)\sqrt{2}, \tag{3}$$
accurate to within a few percent.

The dimensionless quantities $(x_n-1)$, n =1,2, are the basic physical quantities here. They should have a deeper physical meaning than the mass ratios themselves. These quantities measure the deviations from the mass eigenvalue degeneracy, and so they estimate the virtual violation of the lepton mass eigenstate symmetry.

Equation (3) for the lepton mass ratios has two extreme dual solutions with respectively very large and very small violations of the lepton mass symmetry:

(1) A solution with large mass ratios: $x_1 \gg 1$, $x_2 \gg 1$, $x_1 \gg x_2$. Relation (3) shows that if one mass ratio $x_2$ is large, the other one $x_1$ must be much larger. It is appropriate for the CL with $x_1 = m_\mu/m_e$ and $x_2 = m_\tau/m_\mu$, see (1), and can be represented in an exponential form
$$m_\mu/m_e \cong \sqrt{2} \exp \chi, \quad m_\tau/m_\mu \cong \sqrt{2} \exp \chi/2, \quad \chi \gg 1, \tag{1'}$$
with one unknown parameter $\chi$. In this solution, the violation of the mass (lepton flavor) symmetry is a large effect, $\chi \cong \log[(m_\mu/m_e)_{exp}/\sqrt{2}] \cong 4.985$; $m_\mu/m_e \cong \sqrt{2} \exp 5$, $m_\tau/m_\mu \cong \sqrt{2} \exp 5/2$ to within a few percent [1].

(2) A solution with near to unity mass ratios: $x_1 \cong 1$, $x_2 \cong 1$, $(x_1-1) \ll (x_2-1)$. Relation (3) shows that if one mass ratio $x_2$ is close to unity, the other one $x_1$ must be much closer to unity. The violation of the lepton mass symmetry is a small effect here. This other type of solution for the lepton mass ratios can be appropriate only for the neutrinos with a Q-D mass pattern [6],
$$(m_2/m_1) \cong 1, \quad (m_3/m_2) \cong 1, \quad [(m_3/m_2)-1]^2 \cong [(m_2/m_1)-1]\sqrt{2}. \tag{4}$$
It is a probable solution for the neutrinos. With two equations for the atmospheric and solar mass-squared differences and the Eq.(4), there is a full set of three equations for three unknown absolute neutrino masses.

With the definition of the neutrino mass sequence (2), two different cases (A) and (B) are possible for the neutrino solution. Case (A) is as stated in (4). In the other case (B) the ratios ($m_3/m_2$) and ($m_2/m_1$) are interchanged. All estimations below are presented in case (A). They remain intact in case (B).

The neutrino solution (4) can be represented in an exponential form

$$m_3/m_2 \cong \exp(\sqrt{2}\ g^2),\quad m_2/m_1 \cong \exp(\sqrt{2}\ g^4). \tag{5}$$

It contains only one unknown real dimensionless parameter g in the exponents, bound by the consistency condition

$$g^2 \ll 1. \tag{6}$$

The nonlinear relation between the exponents of the two mass ratios in the neutrino solution (5) reflects the nonlinearity feature of the equation (3), unlike the CL solution in (1').

With solution (5), the atmospheric and solar neutrino mass-squared differences are given by

$$\Delta m^2_{atm} = \Delta m_2^2 \equiv (m_3^2 - m_2^2) \cong 2\sqrt{2}\ g^2 m_2^2, \tag{7}$$
$$\Delta m^2_{sol} = \Delta m_1^2 \equiv (m_2^2 - m_1^2) \cong 2\sqrt{2}\ g^4 m_1^2. \tag{8}$$

As a result, it follows

$$m_2^2 \gg \Delta m^2_{atm},\quad m_1^2 \gg \Delta m^2_{sol}, \tag{9}$$
$$\Delta m^2_{atm}/\Delta m^2_{sol} \cong (m_2^2/m_1^2)(1/g^2). \tag{10}$$

Since $(m_2^2/m_1^2) \cong 1$, relation (10) is simplified

$$\Delta m^2_{atm}/\Delta m^2_{sol} \cong 1/g^2. \tag{11}$$

It should be noted, that large ratio of the atmospheric and solar mass-squared differences, $\Delta m^2_{atm}/\Delta m^2_{sol} \gg 1$, is a positive result of the neutrino oscillation experiments [2,3,4]. With (11), this experimental result renders strong evidence in favor of the condition (6) above, and therefore it supports the Q-D neutrino mass ratio pattern (5) and (6).

The absolute neutrino masses follow from (7),(8) and (11):

$$m_2 \cong \sqrt{(\Delta m^2_{atm}/g^2 2\sqrt{2})} \cong \Delta m^2_{atm}/\sqrt{(2\sqrt{2}\ \Delta m^2_{sol})}, \tag{12}$$
$$m_3 \cong m_2 + \Delta m^2_{atm}/2m_2, \tag{13}$$
$$m_1 \cong m_2 - \Delta m^2_{sol}/2m_2. \tag{14}$$

The neutrino mass scale is determined here only by two of the neutrino oscillation data: $\Delta m^2_{atm}$ and $\Delta m^2_{sol}$. Relation (12) can be rewritten in another form

$$(\Delta m^2_{atm}/m_\nu^2)^2 \cong 2\sqrt{2}\ (\Delta m^2_{sol}/m_\nu^2) \tag{12'}$$

where $m_\nu \cong m_2$ is the mass scale of the Q-D neutrinos. The hierarchy of the dimensionless-made mass squared differences of the Q-D neutrinos in (12') conforms to the hierarchy of the CL mass ratios (1).

If supported by data, relation (3) does describe a nonlinear generic feature of the seemingly opposite mass patterns of the neutrinos and CL.

In fact, the statement (12)-(14) for the absolute neutrino masses is a motivated by analogy eigenvalue ansatz for the neutrino mass matrix, to be probed with accurate neutrino mass and oscillation data.

With the best-fit value of the atmospheric neutrino oscillation mass-squared difference [2,7],

$$\Delta m^2_{atm} \cong 2.5 \times 10^{-3}\ eV^2, \tag{15}$$

and the best-fit one for the favored LMA MSW solar neutrino oscillation solution [4,8],

$$\Delta m^2_{sol} \cong 5.5 \times 10^{-5} \text{ eV}^2 , \qquad (16)$$

the ratio in (11) is given by

$$\Delta m^2_{atm}/\Delta m^2_{sol} \cong 45, \ g^2 \cong 1/45. \qquad (17)$$

With another estimation of the best-fit solar neutrino mass-squared difference [9],

$$\Delta m^2_{sol} \cong 7 \times 10^{-5} \text{ eV}^2 , \qquad (18)$$

the ratio in (11) is

$$\Delta m^2_{atm}/\Delta m^2_{sol} \cong 36, \ g^2 \cong 1/36. \qquad (19)$$

The inputs (15) and (16) lead to the estimation of the neutrino mass scale (12),

$$m_2 \cong 0.20 \text{ eV}. \qquad (20)$$

With (15) and (18), the estimation of this scale is

$$m_2 \cong 0.18 \text{ eV}. \qquad (21)$$

With the solar input (18) and the allowed 3σ range from a global analysis [7,9] of the atmospheric neutrino data, instead of (15),

$$1.2 \times 10^{-3} \text{ eV}^2 < \Delta m^2_{atm} < 4.8 \times 10^{-3} \text{ eV}^2, \qquad (22)$$

the estimation for the neutrino mass scale is given by

$$0.09 \text{ eV} < m_2 < 0.34 \text{ eV}. \qquad (23)$$

Though the neutrino mass estimations above are dependent on the exact data values of both the atmospheric and solar neutrino mass-squared differences, they are much more sensitive to the atmospheric data than to the solar ones.

As an important test to date, the considered estimations of the absolute neutrino masses obey the recent cosmological limit $m_\nu < 0.23$ eV (95% C.L.) from the WMAP measurements of cosmic microwave background anisotropy [11], what is a powerful tool for constraining the neutrino mass scale in the Q-D scenario. With neutrino mass scale (12), this cosmological limit on the neutrino mass leads to a restriction,

$$\Delta m^2_{atm}/(2\sqrt{2}\Delta m^2_{sol})^{1/2} < 0.23 \text{ eV}. \qquad (24)$$

According to a subsequent more conservative analysis [12], the restriction is

$$\Delta m^2_{atm}/(2\sqrt{2}\Delta m^2_{sol})^{1/2} < 0.33 \text{ eV}. \qquad (25)$$

These restrictions are compatible with the best-fit values of the atmospheric and solar mass-squared differences in (15), (16) and (18). With the data range (22) for the atmospheric mass-squared difference, a significant inference from the restrictions (24) and (25) is that the LMA MSW solar neutrino oscillation solution is the only one compatible with the present phenomenology of the neutrino mass ratios.

## 3. On a possible indirect evidence of Q-D neutrino masses

In the discussion above, the dimensionless parameter $g^2$ plays a crucial role. It determines the Q-D neutrino mass ratios (5)

and the ratio of the atmospheric neutrino and solar neutrino mass-squared differences (11).

As an interesting coincidence, the value of $g^2$ in (19) is close to the semi-weak coupling constant squared (at $q^2 \cong 0$)
$$g^2 \cong g_W^2/4\pi = G_F\, m_W^2\, \sqrt{2}/\pi \cong 0.034, \qquad (26)$$
$\Delta m^2_{atm}/\Delta m^2_{sol} \cong 30$, $m_\nu \cong 3.26\, \sqrt{(\Delta m^2_{atm})}$. With the input $\Delta m^2_{atm} = (2.0 - 3) \times 10^{-3}$ eV$^2$, it follows $m_\nu \cong (0.15 - 0.18)$ eV and $\Delta m^2_{sol} \cong (6.8 - 10) \times 10^{-5}$ eV$^2$ in agreement with the solar ranges [10,13] $\Delta m^2_{sol} \cong (6 - 9) \times 10^{-5}$ eV$^2$. Since the condition $g^2 \neq 0$ in (5) means neutrino mass splitting, relation (26) hints on a dynamical connection.

To within the same approximation, there is a noticeable relation between the exponents $\chi$ and $g^2$, namely
$$g^2 \cong \chi\, \exp(-\chi) \cong (m_e/m_\mu)\, 5\sqrt{2}, \qquad (27)$$
$$m_3/m_2 \cong \exp(10 m_e/m_\mu), \quad m_2/m_1 \cong \exp[(10 m_e/m_\mu)^2/\sqrt{2}]. \qquad (28)$$

These approximate "coincidental" relations come out into view at the level of exponential lepton mass ratios ($x_n$) in a Q-D neutrino scenario by considering the ratio of the neutrino mass-squared differences in terms of the quantities ($x_n-1$).

Relations (26) and (27) imply a quantitative connection between the close to unity mass ratios of Q-D neutrinos, large mass ratios of the CL, and the weak interactions, though the basic physical meaning of these connections is far beyond the scope of the present low energy phenomenology.

With the new (preliminary) Super-Kamiokande best-fit value (Ref.[10]) for the atmospheric mass-squared difference $\Delta m^2_{atm} = 2.0 \times 10^{-3}$ eV$^2$ in combination with the solar best-fit value (18), the estimation $(\Delta m^2_{atm}/\Delta m^2_{sol}) \cong 29$, instead of (19), is getting closer to the noted relations.

If the inference $g^2 \cong 1/30$ to within a few percent will be supported by further neutrino oscillation data, it will likely mean an indirect evidence of Q-D neutrinos because in that scenario the parameter $g^2$ is properly defined and the large and small quantities $\chi_{exp} \cong 5$ and $g^2_{exp} \cong (\Delta m^2_{sol}/\Delta m^2_{atm})_{exp}$ as exponents determine respectively the CL and neutrino mass ratios. So, in case of Q-D neutrinos, the solar-atmospheric hierarchy parameter from the neutrino oscillation experiments, $(\Delta m^2_{sol}/\Delta m^2_{atm})$, could have a physically meaningful relation to the weak interaction coupling constant (26) and the CL mass-ratio parameter $\chi$ (27).

## 4. Conclusions

An analogy between two basic experimental facts in lepton mass physics — large hierarchy of the CL mass ratios, and large hierarchy of the atmospheric and solar neutrino mass-squared differences — is described by the nonlinear phenomenological equation (3), an extension of the observed relation (1) for the CL data mass ratios. The nonlinearity of this equation is a generic feature of the CL and Q-D neutrino mass-ratio patterns. Two exponential solutions of the equation (3), with large and small exponents, describe respectively the mass ratio patterns of the CL and Q-D neutrinos, Eqs.(1') and (5). Approximate quantitative relations between these exponents are noted. The main results for the absolute neutrino masses are:

(1) The Q-D neutrino mass pattern (5) and (6) is strongly supported by the neutrino oscillation data: $(\Delta m^2_{sol}/\Delta m^2_{atm})_{exp} \ll 1$, this experimental result points to the condition (6) and to the nonlinear connection between the exponents in (5). Three absolute neutrino masses are expressed in terms of two neutrino mass squared differences, as a motivated eigenvalue ansatz for the still unknown exact form of the neutrino mass matrix. The three eigenvalues of the neutrino mass matrix are given in (12), (13) and (14).

(2) The neutrino mass scale (12): $m_\nu \cong \Delta m^2_{atm}/\sqrt{(2\sqrt{2}\Delta m^2_{sol})}$. It is much more sensitive to the atmospheric neutrino data than to the solar ones. By comparison with the available neutrino oscillation data, this neutrino mass scale is located likely within the range $0.1 \div 0.3$ eV, with the best-fit value $m_\nu \cong 0.18 - 0.20$ eV. It should be noted that these estimations of the absolute neutrino masses could be increased in case of a more general Q-D neutrino-CL analogy, but not more than by ~20%, see Appendix.

(3) The estimated neutrino mass scale is compatible with the recent constraints on the absolute neutrino mass from the WMAP data [11,12], with the LMA MSW solution being the only acceptable solar neutrino oscillation solution.

The neutrino mass scale (12) is consistent with the relevant neutrino data to date, and should be confronted with new data. More stringent bounds on the neutrino mass from the coming satellite WMAP measurements (or other relevant data) in combination with more accurate values of $\Delta m^2_{atm}$ and $\Delta m^2_{sol}$ from the neutrino oscillation experiments will probe this neutrino mass scale.

It is noted that indirect evidence in favor of Q-D neutrino masses may come from the neutrino oscillation data on the solar-atmospheric hierarchy parameter $(\Delta m^2_{sol}/\Delta m^2_{atm})$.

I thank A. Habig for the explanation of the new SuperK $\Delta m^2_{atm}$ (preliminary) data.

## Appendix

Consider the known positive result of the neutrino oscillation experiments [2-4]

$$(\Delta m^2_{sol}/\Delta m^2_{atm})_{exp} \equiv r \ll 1. \qquad (A.1)$$

For Q-D neutrinos it follows

$$\Delta m^2_{atm} \cong (x_2^2 - 1)m_2^2, \quad \Delta m^2_{sol} \cong (x_1^2 - 1)m_1^2 \qquad (A.2)$$

and

$$(\Delta m^2_{sol}/\Delta m^2_{atm}) = [(x_1^2 - 1)/(x_2^2 - 1)](m_1^2/m_2^2), \qquad (A.3)$$

and since $m_2^2 \cong m_1^2$ relation (A.3) reads

$$(x_1^2 - 1)/(x_2^2 - 1) \cong r. \qquad (A.4)$$

For an arbitrary Q-D neutrino mass pattern, it should be
$$x_1^2 = \exp \varepsilon_1, \quad x_2^2 = \exp \varepsilon_2, \quad \varepsilon_1 \cong r\varepsilon_2 \ll \varepsilon_2 \ll 1. \tag{A.5}$$
Because of the data condition (A.1), the main terms of expansions in powers of the small parameter $r$ are given by
$$\varepsilon_2 = 2ar^k, \quad \varepsilon_1 \cong 2ar^{k+1}, \tag{A.6}$$
where the unknown power k is arbitrary and the coefficient $a$, $0 < a \ll 1/r$, is not necessarily small. The suggestion above in (3) of a quantitative analogy between the mass-degeneracy-deviation hierarchies of the neutrinos and CL points to the choice k=1 and $a \cong \sqrt{2}$. With the coefficient $a$ less restraint, the mass ratios of Q-D neutrinos may be represented in an exponential form
$$x_2 = \exp ar, \quad x_1 \cong \exp ar^2. \tag{A.7}$$
The physical meaning of the parameter $r \cong g^2$ (A.1) does not depend on the value of the coefficient $a$. The nonlinear relation between the exponents in (A.7) is the same as in (5), the mass ratios (A.7) obey a nonlinear equation
$$(x_2 - 1)^2 \cong a(x_1 - 1), \tag{A.8}$$
which conforms to the nonlinear relation (3).

The condition $a \neq \sqrt{2}$ means that the analogy with CL hierarchy is shifted to a different mass-ratio power. Consider a particular case. The exponential form of the CL mass ratios (1') is a better solution of equation (1) than of Eq.(3). Because of the known large mass ratios of the CL, this difference can be made negligible if the quantitative analogy between neutrino and CL mass-ratio hierarchies is applied to the mass-ratio squared rather than to the mass ratios themselves. The joint phenomenological equation for Q-D neutrino and CL mass-ratio-squared $x_n^2$ reads
$$(x_2^2 - 1)^2 \cong 2(x_1^2 - 1), \tag{A.9}$$
in place of (3). The two exponential solutions of Eq.(A.9) are
$$m_\mu/m_e \cong \sqrt{2} \exp \chi, \quad m_\tau/m_\mu \cong \sqrt{2} \exp \chi/2, \quad \chi_{exp} \cong 5, \tag{A.10}$$
$$m_3/m_2 \cong \exp g^2, \quad m_2/m_1 \cong \exp g^4, \quad g^2_{exp} \cong (\Delta m^2_{sol}/\Delta m^2_{atm})_{exp}, \tag{A.11}$$
for the CL and neutrinos respectively. With regard to the definition (A.7), in (A.11) $a \cong 1$ instead of $a \cong \sqrt{2}$ in (5). With the solution (A.11) instead of (5), the neutrino mass scale $m_\nu$ and all estimations of the absolute neutrino masses of Sec.2 get larger by a factor $2^{1/4}$. It is a ~20% increase, the new estimations remain compatible with the constraints from the WMAP data.

**References**
[1] Particle Data Group, Phys. Rev. **D66**, 010001 (2002).
[2] Super-Kamiokande Collaboration, Y.Fukuda et al., Phys. Lett. **B433**, 9 (1998); **B436**, 33 (1998); Phys. Rev. Lett., **82**, 2644 (1999).
[3] SNO Collaboration, Q.R.Ahmad et al., Phys. Rev. Lett., **89**, 011301 (2002); Phys. Rev. Lett., **87,** 071301 (2001).
[4] KamLand Collaboration, K.Eguchi et al., Phys. Rev. Lett., **90**,

021802 (2003).
- [5] See e.g. Review by R.N.Mohapatra, hep-ph/0211252; and References therein.
- [6] The nearly degenerate neutrino mass pattern was first considered by D.O.Caldwell and R.N.Mohapatra, Phys. Rev. **D48**, 3259 (1993); A.S.Joshipura, Phys. Rev. **D51**.1321 (1995). It is widely discussed in the literature. For a few recent References, V.Barger, S.L.Glashow, D.Marfatia, K.Whisnant, Phys. Lett. **B532**, 15 (2002); Z.Xing, Phys. Rev. **D65**, 077302 (2002); K.S.Babu, E.Ma, J.F.W.Valle, Phys. Lett. **B552**, 207 (2003).
  The recent experimental indications for the neutrinoless double beta decay, H.V.Klapdor-Kleingrothaus et al., Mod.Phys.Lett.**A16**, 2409 (2001), if borne out, can point to a nearly degenerate neutrino mass pattern.
- [7] M.Maltoni, T.Schwetz, M.A.Tortola, J.W.F.Valle, Phys. Rev. **D67**, 013011 (2003); G.Fogli et al., Phys. Rev. **D66**, 093008 (2002).
- [8] SNO Collaboration, Q.R.Ahmad et al., Phys. Rev. Lett. **89**, 011302 (2002); G.L.Fogli et al., Phys.Rev. **D66**, 053010 (2002); V.Barger, D.Marfatia, K.Whisnant, B.P.Wood, Phys. Lett. **B537**, 179 (2002); J.N.Bahcall, M.C.Gonzalez-Garcia, C.Pena-Garay, JHEP 0207 (2002) 057; P.C. de Holanda, A.Yu. Smirnov, Phys. Rev. D66, 113005 (2002); A.Bandyopadhyay et al., Phys. Lett. B540, 14 (20002); P.Aliani et al., Phys. Rev. D67, 013006 (2003); A.Strumia et al., Phys. Lett. B541, 327 (2002).
- [9] V.Barger, D.Marfatia, Phys. Lett. B555(2003)144; G.L.Fogli et al., Phys. Rev. D67(2003)073002; S.Pakvasa, J.W.F.Valle, hep-ph/0301061; M.C.Gonzales-Garcia, C.Pena-Garay, hep-ph/0306001.
- [10] Super-Kamiokande Collaboration, Y.Hayato, talk at HEP 2003, Aachen, Germany, July 17-23, 2003; A.Habig, talk at JCRC 2003, Japan, July 31- Aug.7, 2003, T.Okumura, talk at Int. Workshop, Venice, Italy, Dec., 2003.
- [11] D.N.Spergel et al., Astrophys. J. Suppl., 148 (2003) 175; C.L.Bennett et al., Astrophys. J. Suppl., 148 (2003) 1; G.Hinshaw et al., Astrophys. J. Suppl., 148 (2003) 135; A.Kogut et al., Astrophys. J. Suppl., 148 (2003) 161; A.Pierce, H.Murayama, hep-ph/0302131.
- [12] S.Hannestad, JCAP 05 (2003) 004; astro-ph/0310133;
- [13] V.Barger, D.Marfatia, K.Whisnant, Int.J.Mod.Phys. E12 (2003) 569; A.Y.Smirnov,hep-ph/0311259.